\documentclass[aps,prl,superscriptaddress,showpacs,amsmath,amsfonts,amssymb,showkeys]{revtex4} 
\usepackage[dvips]{graphicx}
\usepackage[latin1]{inputenc}
\usepackage{wrapfig}

\newcommand{\beq}{\begin{equation}} 
\newcommand{\eeq}{\end{equation}}
\newcommand{\bea}{\begin{eqnarray}} 
\newcommand{\eea}{\end{eqnarray}} 
 
\input epsf 
 
\begin{document} 
 
\title{Energy landscape and phase transitions in the
self-gravitating ring model} 
 
\author{Cesare Nardini} 
\affiliation{Dipartimento di Fisica and Centro per lo Studio
delle Dinamiche Complesse, Universit\`a di 
Firenze, via G.~Sansone 1, I-50019 Sesto Fiorentino, Italy}  
\author{Lapo Casetti} 
\email[Corresponding author: ]{lapo.casetti@unifi.it} 
\affiliation{Dipartimento di Fisica and Centro per lo Studio
delle Dinamiche Complesse, Universit\`a di 
Firenze, via G.~Sansone 1, I-50019 Sesto Fiorentino, Italy}  
\affiliation{Istituto Nazionale di Fisica Nucleare, 
Sezione di Firenze, I-50019 Sesto Fiorentino, Italy}  

\date{\today} 
 
\begin{abstract} 
We apply a recently proposed criterion for the existence of phase transitions, which is based on the properties of the saddles of the energy landscape, to a simplified model of a system with gravitational interactions, referred to as the self-gravitating ring model. We show analytically that the criterion correctly singles out the phase transition between a homogeneous and a clustered phase and also suggests the presence of another phase transition, not previously known. On the basis of the properties of the energy landscape we conjecture on the nature of the latter transition. 
\end{abstract} 
 
 
 
\maketitle 

Phase transitions (PTs) are abrupt changes undergone by the collective properties of a system at a
given value of a continuously varying parameter, like temperature or pressure. Common
examples are the melting of ice or the spontaneous magnetization of a ferromagnet. From a
theoretical point of view, PTs are usually  defined as nonanalyticities in thermodynamic
functions\footnote{Other definitions exist, e.g., the non-uniqueness of the Gibbs measure.}.

Our current understanding of many aspects of PTs is remarkable \cite{DombGreen}. This
notwithstanding, there still are some open problems that should not be overlooked. One of such
problems is that we do not know which features of the Hamiltonian of a system induce a PT. The
interaction potential energy $V(q_1,\ldots,q_N)$ among the microscopic degrees of freedom contains,
in principle, all the information on the behavior of a system, at least when the Hamiltonian is 
separable and the kinetic energy term depends only on the momenta, that is for most classical
many-particle systems\footnote{This extends to systems with no kinetic energy, where the
Hamiltonian coincides with the potential energy, like most classical spin models.}. Given $V$, however, we typically do not know what to look for in order to say whether the system has or not a PT. Some {\em
necessary}  conditions on $V$ are known, like those expressed by the Mermin-Wagner theorem
\cite{MerminWagner} excluding PTs for certain classes of  potentials with a continuous symmetry or by
the Van Hove theorem \cite{VanHove} (later generalized by Cuesta and S\'anchez \cite{CuestaSanchez})
excluding PTs for suitable one-dimensional systems, but almost nothing is known about {\em
sufficient} conditions $V$ must fulfill for a PT to take place.

A natural framework for approaching this problem is energy landscape theory \cite{Wales}. The energy
landscape (or more precisely Potential Energy Landscapes, PEL) of a classical system is the graph of the potential
energy function $V: \Gamma_N \mapsto \mathbb{R}$, where $\Gamma_N$ is the configuration space of
coordinates $q_1,\ldots,q_N$, and one searches for those features of the graph which may be
associated with the relevant physical properties of the system. This has been fruitful in the
theory of glasses and disordered systems, whose complex behaviour may be traced back to the
presence of many near-equivalent minima in the PEL \cite{glasses}, as well as in protein folding
theory, where the ability of a protein to fold in a unique native configuration is ascribed to the
overall funnel shape of the PEL \cite{proteins}, with the native state at its bottom. Not only
minima, but also other stationary points of the PEL, i.e., configurations such that $dV = 0$,
play a key role in energy landscape theory and are often referred to as {\em saddles}. Minima (saddles of order zero), or better their basins of attraction, identify the regions (``valleys'') of $\Gamma_N$ which are visited
with a high probability, while saddles of higher order are associated to transition states between
different valleys in the PEL. Saddles of higher energy may become accessible upon increasing the
energy or the temperature, so that different valleys merge and the shape (topology) of the effectively
accessible configuration space changes. When such a change is sufficiently strong, it may entail a
PT; this is the basic idea of the {\em topological hypothesis} about PTs, first suggested in
\cite{CCCP} (see \cite{PhysRep,PettiniBook,KastnerRMP} for reviews). Indeed, Morse theory
\cite{Morse} can be used to show that there is a one-to-one correspondence between saddles of the
PEL and topology changes in the manifolds $M_v = \{ q \in \Gamma_N | V(q) \leq Nv \}$. In addition
to many calculations carried on for model systems (see refs.\ in \cite{PettiniBook,KastnerRMP}), two results have given a
solid basis to the topological approach to PTs. First, Franzosi and Pettini \cite{FranzosiPettini}
proved that under rather general conditions topology changes in the $M_v$'s (thus stationary points
in the PEL) are a {\em necessary} condition for a PT to occur. Second, Kastner, Schnetz and
Schreiber (KSS) \cite{KSS_sing} proved that the microcanonical entropy of a system with a finite
number $N$ of degrees of freedom is nonanalytic at every stationary value of the potential, so that
there is a one-to-one relation between topology changes of the $M_v$'s, stationary points in the
PEL and microcanonical singularities. However,  there are many stationary values of $V$ (their
number being typically proportional to $e^N$); moreover, they usually become dense in all the
interval of values of $V$ as $N\to\infty$ and, in the same limit, the majority (if not all) of the
microcanonical nonanalyticities become irrelevant since they show up as singularities of $N$-th
order derivatives of the entropy \cite{KSS_sing,CasettiKastnerNerattini}. This suggests that the
definition of a PT as a nonanalyticity of a thermodynamic function is unsatisfactory in the
microcanonical ensemble and should be restricted to singularities surviving to the thermodynamic
limit; at the same time, most of the saddles of the PEL (and of the corresponding topology changes of the $M_v$'s) turn out to be unrelated to PTs.

A major step in understanding which are, among all the saddles in the PEL, those that may induce a
nonanalyticity in the entropy which survives in the limit $N\to\infty$ and that may then be related
to thermodynamic PTs was provided again by KSS \cite{KSS_criterion,KSS_sing}. They put forward a
discriminating criterion whose content is essentially that  such saddles must become sufficiently
``flat'' as $N\to\infty$. 
The KSS criterion, as well as most (if not all) of the other tools proposed within the topological approach to PTs, has been successfully applied \cite{KSS_criterion,KSS_sing} until now only to models that are $(i)$ exactly solvable and $(ii)$ whose energy landscape is completely characterized, i.e., essentially all the stationary points and stationary values can be analytically calculated. Such models, although clearly important as a conceptual benchmark, are highly nongeneric. 

The aim of the present Letter is to analyze further the potentialities of the KSS criterion by applying it to a model system that, contrary to those previously studied, is {\em not} exactly solvable and whose energy landscape is {\em not} completely characterized analytically. Nonetheless, the criterion is able to single out the stationary points associated to a PT; it correctly detects the homogeneous-clustered PT and also suggests the possibility of another PT, previoulsy unknown. Moreover, in spite of the fact that the model is not exactly solvable so that all the information on its equilibrium behavior must be obtained numerically, our analysis has been carried out analytically. Our results then show that the KSS criterion can be effectively applied to generic systems, more complex than those previously addressed in the topological approach to PTs, and that it may even provide new insights and exact results otherwise unattainable.

{\em The KSS criterion.} Let us briefly discuss the KSS criterion (for details see \cite{KSS_sing}). Consider a classical Hamiltonian of the form
\beq
{\cal H}(p,q) = \frac{1}{2}\sum_{i = 1}^N p_i^2 + V(q_1,\ldots,q_N)
\label{H}
\eeq
and assume that the stationary points of $V$ are isolated and that their overall number grows at most exponentially\footnote{This is the typical case; a proof for a special case can be found in \protect\cite{Schilling}.} with $N$. Then the microcanonical entropy per degree of freedom $s(\varepsilon)$ can be nonanalytic in the limit $N\to\infty$ at $\varepsilon = \varepsilon_c$ only if the following two conditions are met. First, there is a sequence $\left\{q^N_c\right\}_{N=1}^\infty$ of stationary points of $V$ whose corresponding stationary values converge to $v_c = \langle v \rangle (\varepsilon_c)$, where $v =  V/N$ and $\langle v \rangle$ denotes its statistical expectation value, i.e.,
\beq
\lim_{N\to\infty} v\left(q^N_c \right) = v_c \,  .
\label{lim_cond}
\eeq
Second, the Hessian matrix $\mathbb{H}_V$ of the potential $V$ computed on the stationary configurations $q_c^N$ is such that 
\beq
\lim_{N\to\infty} \left| \det \left[\mathbb{H}_V \left(q^N_c \right) \right] \right|^{1/N} = 0\, \, .
\label{geom_cond}
\eeq
Since the eigenvalues of $\mathbb{H}_V$ can be seen as curvatures of the PEL, Eq.\ (\ref{geom_cond}) means that the saddles become asymptotically ``flat''. We note that to check whether 
Eqs.\ (\ref{lim_cond}) and (\ref{geom_cond}) are satisfied we do not need to know {\em all} the saddles of the PEL; it is sufficient to determine the ``right'' ones. This is a big difference with respect to other criteria previously proposed, which made use of topological invariants like the Euler characteristic whose calculation requires to know all the saddles of the PEL (and their order) \cite{PhysRep,KastnerRMP}.

{\em The model.} The model we studied is the Self-Gravitating Ring (SGR), first introduced in \cite{SGR1} as a simplified model of a self-gravitating system. It is a model of $N$ points of unitary mass moving on a circle of unitary radius and mutually interacting via gravitational forces, regularized at short distances. Its Hamiltonian is of the form (\ref{H}) with potential
\beq
V = - \frac{1}{2N\sqrt{2}}\sum_{i,j = 1}^N \frac{1}{\sqrt{1 - \cos\left(q_i - q_j \right) + \alpha}}\,\, ,
\label{VSGR}
\eeq
where $q_i \in (-\pi,\pi]$, $i = 1,\ldots,N$, are the angles giving the position of the $i$-th particle on the ring and $\alpha > 0$ is the softening parameter regularizing the potential for $(q_i - q_j) \to 0$. The $\frac{1}{N}$ factor in Eq.\ (\ref{VSGR}) ensures extensivity according to the Kac prescription, although the model is non additive, the interactions being long-range \cite{longrangereview}.

Numerical results ($N$-body simulations \cite{SGR1} as well as variational calculations in the $N\to\infty$ limit \cite{SGR2}) for the microcanonical thermodynamics show that the system has a PT separating a homogeneous high-energy phase from a clustered phase where the rotational symmetry of the potential is spontaneously broken. If the softening parameter $\alpha$ is sufficiently small the clustered phase contains an energy interval with negative specific heat, there is nonequivalence between the microcanonical and the canonical ensemble and the qualitative behavior of the system is very similar to that of three-dimensional self-gravitating models. In the opposite limit $\alpha \to \infty$ SGR becomes equivalent to the ferromagnetic HMF model \cite{HMF}. Assuming that in the high-energy phase the particles are homogeneously distributed on the ring, as suggested by the numerical results, one can calculate the expectation value of the potential energy density at the transition in the limit $N \to \infty$, obtaining
\beq
v_c = \langle v \rangle(\varepsilon_c) = \frac{1}{\pi\sqrt{2(2 + \alpha)}}\, {\cal K}\left( \frac{2}{2 + \alpha}  \right) \, ,
\label{vc}
\eeq
where ${\cal K}(x) = \int_0^{\pi/2} {d\vartheta}/{\sqrt{1 -x \sin^2 \vartheta}}$ is the complete elliptic integral of the first kind. 

 {\em Saddles of the landscape and PTs in the SGR model.} Let us now study the PEL of the SGR. First of all let us note that the stationary points of the potential (\ref{VSGR}) are not isolated, due to its rotational invariance. However, this difficulty can be circumvented by fixing the value of one of the $q$'s, which has an irrelevant effect on the thermodynamic functions as $N\to\infty$ and only fixes the position of the center of the cluster in the broken-symmetry phase: from now on we assume $q_1 \equiv 0$. The stationary points of $V$ are the solutions of the form $(\bar q_1 \equiv 0,\bar q_2,\ldots,\bar q_N)$ of the $N$ coupled nonlinear equations $\nabla V = 0$, i.e.,
\beq
\frac{1}{2 N \sqrt{2}} \sum_{i = 1}^N \frac{\sin \left(q_i - q_k \right)}{\left[1 - \cos\left(q_i - q_k \right) + \alpha\right]^{3/2}} = 0\,\, , 
\label{nablaVSGR}
\eeq
with $k = 1,\ldots,N$. Physically the above equations mean that the force acting on each particle is radial. There are at least two classes of solutions that can be easily found. First of all there are the solutions we shall refer to as {\em $0$-$\pi$ saddles}, where $N_\pi$ particles are in $q = \pi$ and the others are in $q = 0$, with $0 \leq N_\pi \leq N-1$. We shall denote such configurations as $q_{n_\pi}$, where $n_\pi = N_\pi/N$. The other class of solutions of Eqs.\ (\ref{nablaVSGR}) that is easily found is that of configurations where the particles are in the $p$ vertices of a regular polygon, with the same number $r$ of particles in each vertex; from symmetry considerations one sees that the force can be only radial. We shall refer to the latter solutions as {\em polygonal saddles} and we shall denote them as $q_{p,r}$, with $N = pr$. There are also many other saddles of the SGR potential that do not belong to any of these two classes. For instance, consider three angles $0$, $\gamma$ and $\delta$, and put $N_0$ particles in $0$, $N_\gamma$ in $\gamma$ and $N_\delta$ in $\delta$. It can be shown that 
for almost any value of $\gamma$ and $\delta$ such that $0< \gamma< \pi$, $-\pi < \delta < 0$ and $0<\delta - \gamma < \pi$ one can choose sufficiently large $N$, $N_0$, $N_\gamma$ and $N_\delta$ such that this configuration is stationary; 
moreover, one can find arguments suggesting that also more complex stationary configurations exist \cite{tesicesare}.  The polygonal and $0$-$\pi$ saddles, then, are not the only saddles of the PEL of the SGR model. 

This does not prevent the KSS criterion to be effectively applied to this model. Let us calculate the stationary values $v = V/N$ corresponding to the $0$-$\pi$ and polygonal saddles, respectively. For $q_{n_\pi}$ we have
\beq
v(n_\pi) = - \frac{1}{2 \sqrt{2}} \left[\frac{\left(1 - n_\pi \right)^2 + n_\pi^2}{\sqrt{\alpha}} + \frac{2 n_\pi\left(1 - n_\pi \right)}{\sqrt{2 + \alpha}} \right] \, .
\label{v(npi)}
\eeq
The maximum of $v(n_\pi)$ is attained for $n_\pi = 1/2$ and the minimum, corresponding to $n_\pi = 0$, is also the absolute minimum of the potential. Hence
\beq
- \frac{1}{2 \sqrt{2\alpha}} \leq v(n_\pi) \leq - \frac{1}{4 \sqrt{2}} \left[\frac{1}{\sqrt{\alpha}} + \frac{1}{\sqrt{2 + \alpha}} \right] \, \, ,   
\eeq
and the values of $v(n_\pi)$ become dense in the above interval as $N\to\infty$. For $q_{p,r}$ we have that the stationary values depend only on the number of vertices $p$:
\beq
v(p) = - \frac{1}{2 p \sqrt{2}} \sum_{j = 0}^{p-1}\frac{1}{\sqrt{1 -\cos\left(\frac{2\pi j}{p} \right) + \alpha}}\,\, .   
\eeq
The function $v(p)$ is monotonously increasing with $p$ and 
\beq
v(p) \geq - \frac{1}{4 \sqrt{2}} \left[\frac{1}{\sqrt{\alpha}} + \frac{1}{\sqrt{2 + \alpha}} \right] \, \, ,   
\eeq
so that $v(p) > v(n_\pi)$ $\forall\, p,n_\pi$. Moreover, a simple calculation shows that $\lim_{p\to\infty} v(p) = v_c$, where $v_c$ is given by Eq.\ (\ref{vc}) and is also the upper bound of the potential energy per particle. We have thus two nontrivial results: $(i)$ since as $N\to\infty$ the distance between two successive stationary values tends to zero in both cases of $v(n_\pi)$ and $v(p)$, although $0$-$\pi$ and polygonal saddles are not the only stationary values of $V$ their stationary values do encompass all the available values of the potential energy of the model; $(ii)$ the sequence of stationary values $v(p)$ converges to the critical potential energy $v_c$ of the PT between the homogeneous and the clustered phase as $p \to\infty$. 

The last result suggests to investigate whether the KSS criterion is satisfied for $v = v_c$, i.e., whether a sequence of saddles satisfying Eqs.\ (\ref{lim_cond}) and (\ref{geom_cond}) exists, with $v_c$ given by Eq.\ (\ref{vc}). To this end, consider the sequence of polygonal saddles with one particle in each vertex, $q_{N,1}$, where $N$ is prime; Eq.\ (\ref{lim_cond}) is satisfied---see item $(ii)$ above. It remains to show that 
also Eq.\ (\ref{geom_cond}) is satisfied. Define
\beq
f_k = \frac{1}{4N\sqrt{2}} \frac{2 - (2 + 2\alpha)\cos\frac{2\pi k}{N} + \sin^2\frac{2\pi k}{N}}{\left( 1 - \cos\frac{2\pi k}{N} + \alpha\right)^{5/2}}~;
\eeq
then the diagonal elements of the Hessian matrix are $
\left[\mathbb{H}_V\left(q_{N,1}\right)\right]_{kk}  =  - \sum_{i = 0}^{N-1} f_k$  
and the off-diagonal ones are $ 
\left[\mathbb{H}_V\left(q_{N,1}\right)\right]_{lk}  =  f_{k - l}$, 
so that the Hessian calculated in $q_{N,1}$ is a circulant matrix. Using the Hadamard inequality \cite{Hadamard} to obtain an upper bound to the absolute value of the 
determinant of a matrix as the product of the Euclidean norms of its rows 
and observing that in a circulant matrix all the rows have the same norm, after some algebra we can write 
\beq
\lim_{N\to\infty} \left| \det \left[\mathbb{H}_V \left(q_{N,1} \right) \right] \right|^{1/N} \leq \lim_{N\to\infty} \left| \sum_{k=0}^{N-1} f_k \right|\,.  
\label{hadamard_estimate}
\eeq
The r.h.s.\ of the above equation can be written as an integral, so that  Eq.\ (\ref{geom_cond}) is satisfied if 
\beq
\int_0^{2\pi} dx\, \frac{2 - (2+2\alpha)\cos x + \sin^2 x}{\left(1 - \cos x + \alpha \right)^{5/2}}  =  0\, ;
\eeq
the above result can be proved by explicit integration, so that the KSS criterion is satisfied for $v = v_c$. 

The properties of the circulant matrices \cite{Davis} 
can also be used to prove that Eq.\ (\ref{geom_cond}) is {\em not} satisfied for all the stationary values $v(p) \not = v_c$ (details may be found in \cite{tesicesare}). Hence, as far as the stationary values of $V$ associated to polygonal saddles are concerned, we have shown analytically that the KSS criterion is satisfied if and only if $v = v_c$. Since the system is not exactly solvable, this is, to the best of our knowledge, the only analytical indication of the presence of a PT between a homogeneous and a clustered phase in the SGR model.

To complete the analysis of the whole range of potential energy values we now turn to the $0$-$\pi$ saddles. Since the stationary values (\ref{v(npi)}) depend only on $n_\pi = N_\pi/N$, let us consider sequences of saddles $q_{n_\pi}$ with varying $N$ at fixed $n_\pi$. The Hessian calculated on such saddles can be written as $\mathbb{H}_V \left(q_{n_\pi} \right) = \mathbb{D} + \mathbb{A}$, where $\mathbb{A}$ has rank 2 and $\mathbb{D} = \text{diag}(d_1,\ldots,d_k)$ with
\beq
d_k = \left\{
\begin{array}{cl}
\frac{N_\pi + 1}{N \lambda_1} - \frac{N-N_\pi}{N\lambda_2} & ~ \text{if}~ 1 \leq k \leq N_\pi\, ,\\
\frac{N-N_\pi + 1}{N \lambda_1} -\frac{N_\pi}{N\lambda_2} & ~ \text{if}~ N_\pi + 1 \leq k \leq N\, ;
\end{array} \right.
\eeq 
here $\lambda_1 = 2\sqrt{2\alpha^3}$ and $\lambda_2 = 2\sqrt{2(2 +\alpha)^3}$. Using such decomposition one can prove that only the elements of $\mathbb{D}$ contribute to $\left|\det\left[\mathbb{H}_V \left(q_{n_\pi} \right)\right]\right|^{1/N}$ when $N\to\infty$: 
\beq
\lim_{N\to\infty} \left|\det\left[\mathbb{H}_V \left(q_{n_\pi} \right)\right]\right|^{1/N} = \left[a(n_\pi)\right]^{n_\pi} \left[b(n_\pi)\right]^{1-n_\pi} ~,   
\label{detzeropi}
\eeq
where $a(n_\pi) = n_\pi/ \lambda_1- (1 - n_\pi)/\lambda_2$ and $b(n_\pi) = (1 - n_\pi)/\lambda_1- n_\pi /\lambda_2$. The quantity in Eq.\ (\ref{detzeropi}) vanishes if and only if $n_\pi = n^c_\pi$ or $n_\pi = 1 - n^c_\pi$, where $n^c_\pi = \alpha^{3/2}/\left[(2+\alpha)^{3/2} + \alpha^{3/2}\right]$
and $v(n_\pi^c) = v(1-n_\pi^c) = v_c'$, with
\beq
v_c' = -\frac{4 + \alpha[6 + \alpha(5+2\alpha)]}{\sqrt{2\alpha}\left[(2+\alpha)^{3/2} + \alpha^{3/2}\right]}~.
\label{vcp}
\eeq
The KSS criterion is then satisfied not only at $v_c$ given by Eq.\ (\ref{vc}) but also at $v_c'$ given by Eq.\ (\ref{vcp}). Is there a PT in the SGR model at $\varepsilon_c'$ such that $\langle v \rangle(\varepsilon_c') = v'_c$, as suggested by the KSS criterion? We do not have a final answer yet. Previous numerical studies of the SGR model \cite{SGR1,SGR2} have not detected such a PT. However, we note that $v_c'$ is extremely close to the absolute minimum of the potential for small values of $\alpha$, where the SGR model is close to a ``true'' self-gravitating system. Conversely, $v_c' \simeq v_c$ for large values of $\alpha$, where it behaves like a mean-field ferromagnet. Only for $\alpha \simeq 1$ one has that $v'_c$ is clearly separated from both the mimimum and $v_c$, but these values of $\alpha$ have not been thoroughly studied in \cite{SGR1,SGR2}. Moreover, the analysis of a related model \cite{tesicesare} suggest that such a PT may indeed exist but its effect on the thermodynamic quantities may be weak, so that it may not be easy to detect numerically. If such a PT exists, which phases does it separate? An answer may be suggested again by an analysis of the PEL. Computing the number of negative eigenvalues of the Hessian one can show that saddles $q_{n_\pi}$ with $n_\pi < n_\pi^c$ are proper saddles, i.e., saddles of order greater than zero, while saddles with $n_\pi > n_\pi^c$ are minima. Minima are visited with high probability, at variance with higher-order saddles. Hence we conjecture that for $\langle v \rangle < v'_c$ the equilibrium phase is such that the fraction of particles that may cross $q = \pi$, i.e., visit the whole circle, is zero, while it becomes nonzero as $\langle v \rangle > v'_c$, i.e., at $\varepsilon > \varepsilon'_c$.

{\em Summary.} We have applied a recently proposed criterion for PTs, referred to as the KSS criterion, to the SGR model which, at variance with the few models the criterion had previously been applied to, is not exactly solvable. We have shown analytically that the criterion correctly singles out the PT between a homogeneous and a clustered phase. We have also shown that the criterion indicates the possible presence of another PT, not previously known, and we have conjectured on its nature. Our results suggest that the criterion may be effectively applied to even more complex systems, if one can guess which stationary configurations can be related to PTs.

\end{document}